\documentclass[runningheads]{llncs}
\usepackage{graphicx}
\usepackage{float}
\usepackage{setspace}
\usepackage{multirow}
\usepackage{enumitem}
\usepackage{cite}
\usepackage{xcolor}
\usepackage{array}
\usepackage[caption=false]{subfig}
\usepackage{colortbl}
\usepackage{hhline}
\usepackage{makecell}

\begin{document}

\title{CyberSecurity Challenges: Serious Games for Awareness Training in Industrial Environments}

\author{
  Tiago Gasiba\inst{1,2} \and
  Ulrike Lechner\inst{2}  \and
  Maria Pinto-Albuquerque\inst{3}
}
\authorrunning{T. Gasiba et al.}
\institute{
  Siemens AG, Munich, Germany\\
  \email{tiago.gasiba@siemens.com}
  \and
  Universität der Bundeswehr München, Munich, Germany\\
  \email{tiago.gasiba@unibw.de} \email{ulrike.lechner@unibw.de}
  \and
  Instituto Universitário de Lisboa (ISCTE-IUL), ISTAR, Lisboa, Portugal\\
  \email{maria.albuquerque@iscte-iul.pt}
}

\maketitle

\begin{abstract}
Awareness of cybersecurity topics, e.g., related to secure coding guidelines, enables software developers to write secure code. This awareness is vital in industrial environments for the products and services in critical infrastructures. In this work, we introduce and discuss a new serious game designed for software developers in the industry. This game addresses software developers' needs and is shown to be well suited for raising secure coding awareness of software developers in the industry. Our work results from the experience of the authors gained in conducting more than ten CyberSecurity Challenges in the industry. The presented game design, which is shown to be well accepted by software developers, is a novel alternative to traditional classroom training. We hope to make a positive impact in the industry by improving the cybersecurity of products at their early production stages.

\keywords{
IT Security
\and Cybersecurity
\and Awareness
\and Secure Software Development
\and Industry
\and Critical Infrastructures
\and Serious Game
}
\end{abstract}

\section{Introduction}
If not addressed during the early stages of software design and implementation, software development errors and security vulnerabilities can end up in a final product or service. Security vulnerabilities can result in serious negative consequences for society, the customer, and the company that produced the software. Think, e.g., of critical infrastructures as the grid, transportation, or production lines: a security vulnerability in the code may cause interruptions in service quality for individual customers when critical machinery or information systems fail or even for society when critical infrastructure fails. Over the last years, the number of industrial security-related incidents has been increasing, which has resulted in severe incidents, leading to a substantial financial impact, reaching up to 1.6\% of GDP in some EU countries \cite{ENISA2016}.

To address these issues, products and services provided by the industry must follow IT security standards. These standards mandate the implementation of a secure software development lifecycle and secure coding guidelines that must be followed to write secure code. Prominent examples of these standards for industrial environments are the IEC 62443 \cite{2018_62443_4_1}, ISO 27001 \cite{ISO2013a}, and the Grundschutzkatalog from the Bundesamt für Sicherheit in der Informationstechnik (BSI) \cite{2016_Grundschutz_Katalog}. Examples of secure coding guidelines widely used in the industry are the SEI-CERT Java Secure Coding Guidelines and SEI-CERT C/C++ Secure Coding Guidelines, both from Carnegie Mellon \cite{UniCarnegieMellon2019}. The Open Web Application Security Project (OWASP, \cite{OWASP}) and the BSI (BSI 5.21, \cite{BSI2014}) provide secure coding guidelines which are specific for web application development and widely used in the industry.

These standards provide a much-needed basis that establishes ground rules required to produce secure products and services. The effectiveness of these standards is related to the level of awareness and understanding of the standards by the persons directly affected by them: software developers. However, a recent study by Patel et al. \cite{gitlab_2019} has shown that more than 50\% of software developers cannot spot software vulnerabilities in source code. This lack of awareness about secure coding is a problem that needs to be addressed.

Among others, a possible way to address this issue is to provide training to software developers on secure coding. We present a new serious games designed to raise awareness and train software developers in secure coding in this work. The serious game, named CyberSecurity Challenges, is an adaption of the capture-the-flag game genre. Capture-the-flag was initially developed in the penetration testing community to practice and train offensive IT-security skills. The idea is that by attacking a system, well-trained penetration testers can discover vulnerabilities in products and services that can be fixed before final shipment to the customer. However, since these activities require a full or partially developed project, they often occur late in the software development stages. We propose using an adapted version of the game, which targets software developers, focuses on the defensive perspective, and has the primary goal of increasing awareness of secure coding guidelines and secure coding best practices. Furthermore, we show how our concept can be used for onsite IT-Security Awareness Workshops and how it can be adapted for online training.

This work is organized as follows: in section 2, the authors briefly discuss previous work related to the cybersecurity challenges. Section 3 introduces the CyberSecurity Challenges and discusses challenges based on open-source components and the Sifu platform.
Section 4 discusses the games' evaluation in an industrial context through survey results, participant feedback, and lessons learned.
Finally, section 5 summarizes and concludes the paper.

\section{Related Work}
Although several methods exist to deal with software vulnerabilities, e.g., requirements engineering and code reviews, we focus on awareness training for software developers.
Several previous studies indicate that software developers lack secure programming awareness and skills~\cite{Assal2019,gitlab_2019,tahaei2019survey}.
In 2020, Bruce Schneier, a well-known security researcher, and evangelist stated that {\it less than 50\% of software developers can spot security vulnerabilities in software} \cite{Schneier2020}. His comment adds to a discussion on secure coding skills: In 2011, Xie et al.~\cite{Xie2011} did several interviews with 15 senior professional software developers in the industry with an average of 12 years of experience. Their study has shown a disconnect between software security concepts and their role in their jobs.
Awareness training on Information security is addressed in McIlwraith \cite{McIlwraith2006}, which provides a systematic methodology and a baseline for implementing awareness training.

There is a stream of literature on compliance with security policies, which deals with general employees, not with software developers specifically. This stream of literature explores many reasons why people do not comply with IT-security policies. The unified framework by Moody et al.~\cite{moody2018toward} summarizes the academic discussion on compliance with IT-security policies. Empirical findings conclude that neither deterrence nor punishment, such as, e.g., public blame, works to increase compliance. However, increasing IT-security awareness increases the level of compliance \cite{Stewart2012}. In their seminal review article, Hänsch et al.~\cite{2014_Benenson_Defining_Security_Awareness} define IT-security awareness in the three dimensions: {\it Perception}, {\it Protection}, and {\it Behavior}. The concept of IT-security awareness is typically used in IT security management contexts. We adapt these concepts to software developers as follows \cite{gasiba_wi}: perception - knowledge of existing software vulnerabilities, protection - knowing the existing mechanisms, e.g., secure coding guidelines and software development best practices, that avoid software vulnerabilities, and behavior - knowledge and intention to write secure code.

Graziotin et al.~\cite{Graziotin2018} show that \textit{happy developers are better coders}, i.e., produce higher quality code and software.
Their work suggests that by keeping developers happy, we can expect that the code they write has a better quality and, by implication, be more secure.
Davis et al.\cite{Davis2014} show, in their construct, that cybersecurity games have the potential to increase the overall happiness of software developers. Their conclusions support our approach to use a serious game to train software developers in secure coding.
Awareness games are a well-established instrument in information security. They are discussed in de-facto standards as the BSI Grundschutz-Katalog \cite{2016_Grundschutz_Katalog} (M 3.47, Planspiele) as one means to raise awareness and increase the level of security. Frey et al.~\cite{Maria2019} show both the potential impact of playing cybersecurity games on the participants and show the importance of playing games as a means of cybersecurity awareness. They conclude that cybersecurity games can be a useful means to build a common understanding of security issues. Rieb et al. \cite{Rieb2018} provide a review of serious games in cybersecurity and conclude that there are many approaches. The games listed mainly address information security rather than secure coding. Documented and evaluated games are \cite{2016_Beckers_Serious_Game} and \cite{Rieb2018}.

Capture-the-flag is one particular genre of serious games in the domain of Cybersecurity \cite{Davis2014}. Game participants win flags when they manage to solve a task. Forensics, cryptography, and penetration testings are skills necessary for solving tasks and capturing flags.
The present work uses serious games to achieve the goal of {\it raising secure coding awareness of software developers in the industry}. 
Previous work on selected design aspects and a smaller empirical basis on the CSC includes \cite{Gasiba2020_RankingSCG,Gasiba2020_PlayerProfile,gasiba_re19,Gasiba2020_CyberICPS,Gasiba2020_CyberICPS_Journal,Gasiba2020_QUATIC,Gasiba2020_TrustCOMM}.

\section{CyberSecurity Challenges}
In this section, we introduce the CyberSecurity Challenges (CSC), which were developed to raise awareness on secure coding. We also present a detailed discussion on creating these games (1) by using existing open-source components, and (2) using the open-source Cybersecurity Challenge platform developed by the authors - the Sifu platform.

\subsection{What are CyberSecurity Challenges}

\begin{figure}[http]
    \centering
    \includegraphics[width=\columnwidth]{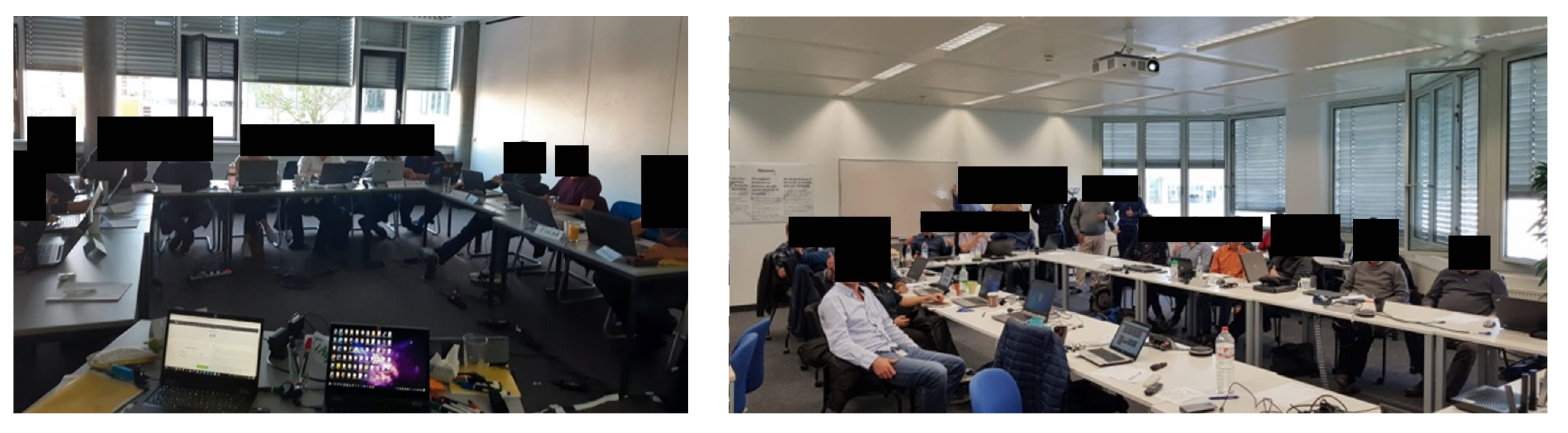}
    \caption{CyberSecurity Events - On-site Events}
    \label{fig:teams}
\end{figure}

CyberSecurity Challenges (CSC) are a genre of serious games developed with the specific purpose of raising awareness of industrial software developers in the topic of secure coding and secure coding guidelines.
Figure \ref{fig:teams} shows two examples of CSC events in the industry.

\begin{figure}[http]
    \centering
    \includegraphics[width=.75\columnwidth]{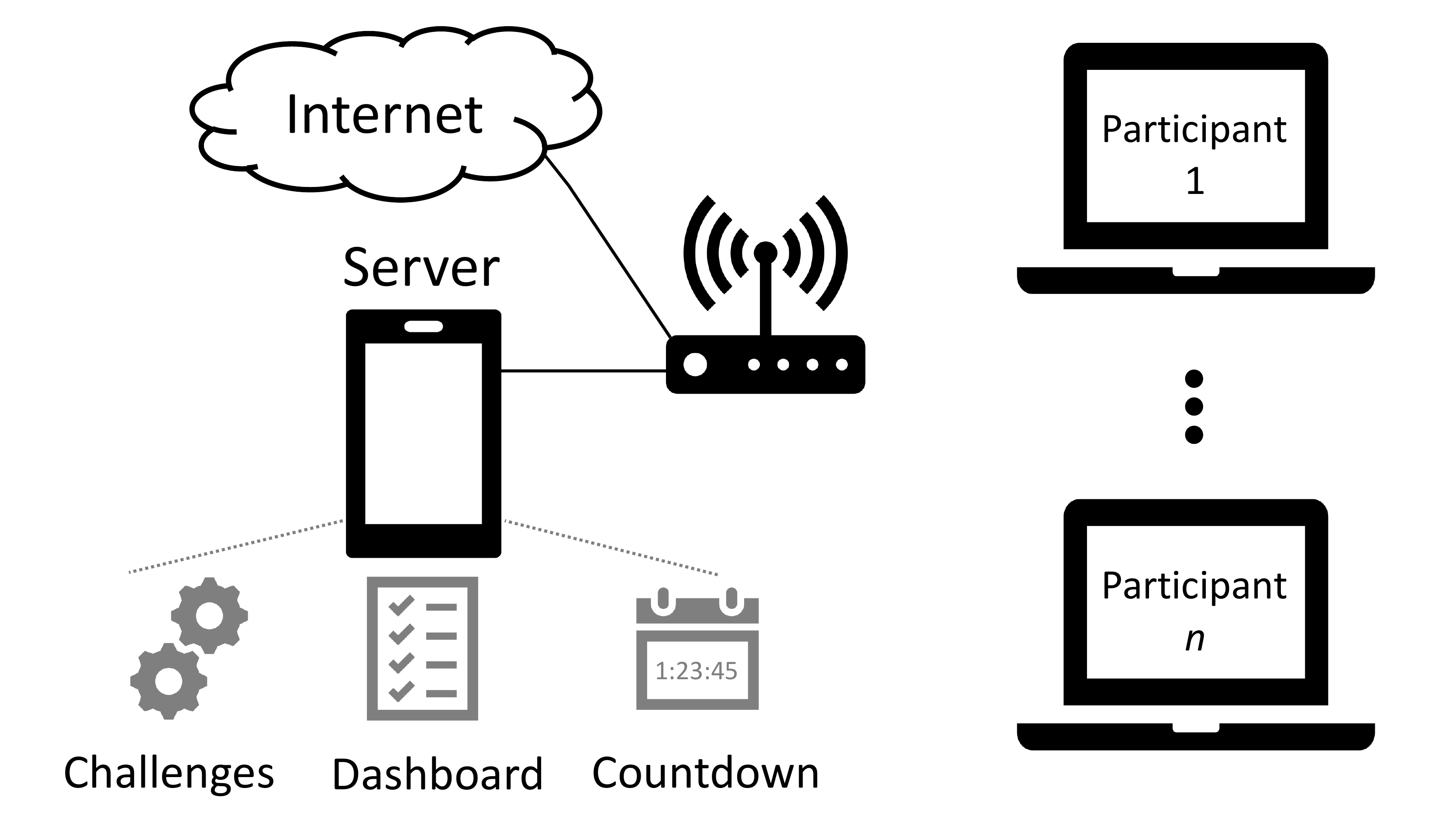}
    \caption{Architecture of CyberSecurity Challenges infrastructure}
    \label{fig:CSC_Architecture}
\end{figure}

The game consists of a platform where several participants (i.e., software developers) form teams that compete against each other in solving secure coding challenges. The challenges consist of exercises that are developed primarily to address software development vulnerabilities. Solving the challenges requires the participants to know and follow secure coding guidelines.
Figure \ref{fig:CSC_Architecture} depicts the general architecture of CyberSecurity Challenges (CSC), which consists of the following components: Challenges, Dashboard, and Countdown. 

The challenges represent the individual exercises that the participants must solve to gain points.
The dashboard displays the available challenges and is used to control each team's current status regarding the number of gathered points. Figure \ref{fig:dashboard} shows an example of a dashboard based on the open-source CTFd platform. Upon solving a challenge, the participants receive a flag. This flag is represented by a random-like string that can be redeemed for points in the dashboard. The reward on the number of points is related to the difficulty level of the challenge. 
The countdown component consists of a timer that, when expired, automatically locks the dashboard, preventing further submission of flags. The countdown timer is also used to incentivize the competitiveness of the players on solving the challenges. One or more coaches take part in the game by aiding every team and every participant during the gameplay, such that no one gets stuck or lost while solving the exercises. The coaches also supervise the gameplay to ensure that the desired game objectives, e.g., learning goals, are achieved.
In the end, the team with the highest amount of points wins the challenge. Nevertheless, all teams and players are winners since, by participating in the game, awareness of secure coding is stimulated. The game's competitive nature increases the fun, contributes to the overall awareness level of every player, and ensures a memorable event that can have long-lasting impressions.
 
The different CSC challenges can be implemented in two ways: 1) using open-source components or 2) using self-developed components. In the first case, the challenges are implemented through adaptation, re-use, and re-purposing existing open-source projects and components. This method's main advantage is the reduced cost of implementation of individual challenges while outsourcing their maintenance. In the second case, the challenges can be better adapted to internal company policies while also focusing more on the defensive perspective.
The architecture shown in Figure \ref{fig:CSC_Architecture} was initially developed for onsite events. A recent installment of the game \cite{Gasiba2020_CyberICPS_Journal} allows the game not only to be played remotely but also to include an intelligent coach based on artificial intelligence techniques.
In the following, we present a more detailed introduction of the CSC game implementation based on open-source components and the Sifu platform.

\subsection{CyberSecurity Game}
The CSC game was developed in the industry, focusing on Web and C/C++ developers.
In contrast to C/C++, for the web challenges, it was decided not to focus on a single programming language or framework since many of these programming languages and frameworks are in everyday use in the company where the CSC game was developed. In this case, we chose a generic approach based on the Open Web Application Security Project - OWASP~\cite{OWASP}.
The challenges' design took two approaches: 1) based on open-source components and 2) design of own challenges.
A common approach to the design of the challenges is given in~\cite{Gasiba2020_QUATIC}.
Each challenge is presented to the participants according to the following phases: {\it Phase 1} - introduction, {\it Phase 2} - challenge, and {\it Phase 3} - conclusion.
Phase 1 presents an introduction to the challenge and sets up the scenario; the core part of the challenge is phase 2; phase 3 concludes the challenge by adding additional text related to secure coding guidelines or further questions related to phase 2.
The types of challenges are Single-Choice Questions, Multiple-Choice Questions, Text-Entry Questions, Associate-Left-Right, Code-Snippet Challenge, and Code-Entry Challenge.

\subsubsection{Challenges using Open-Source Components}
\begin{figure}[ht]
    \centering
    %\vspace*{-1em}
    \begin{minipage}{.49\textwidth}
        \centering
        \includegraphics[width=.99\linewidth]{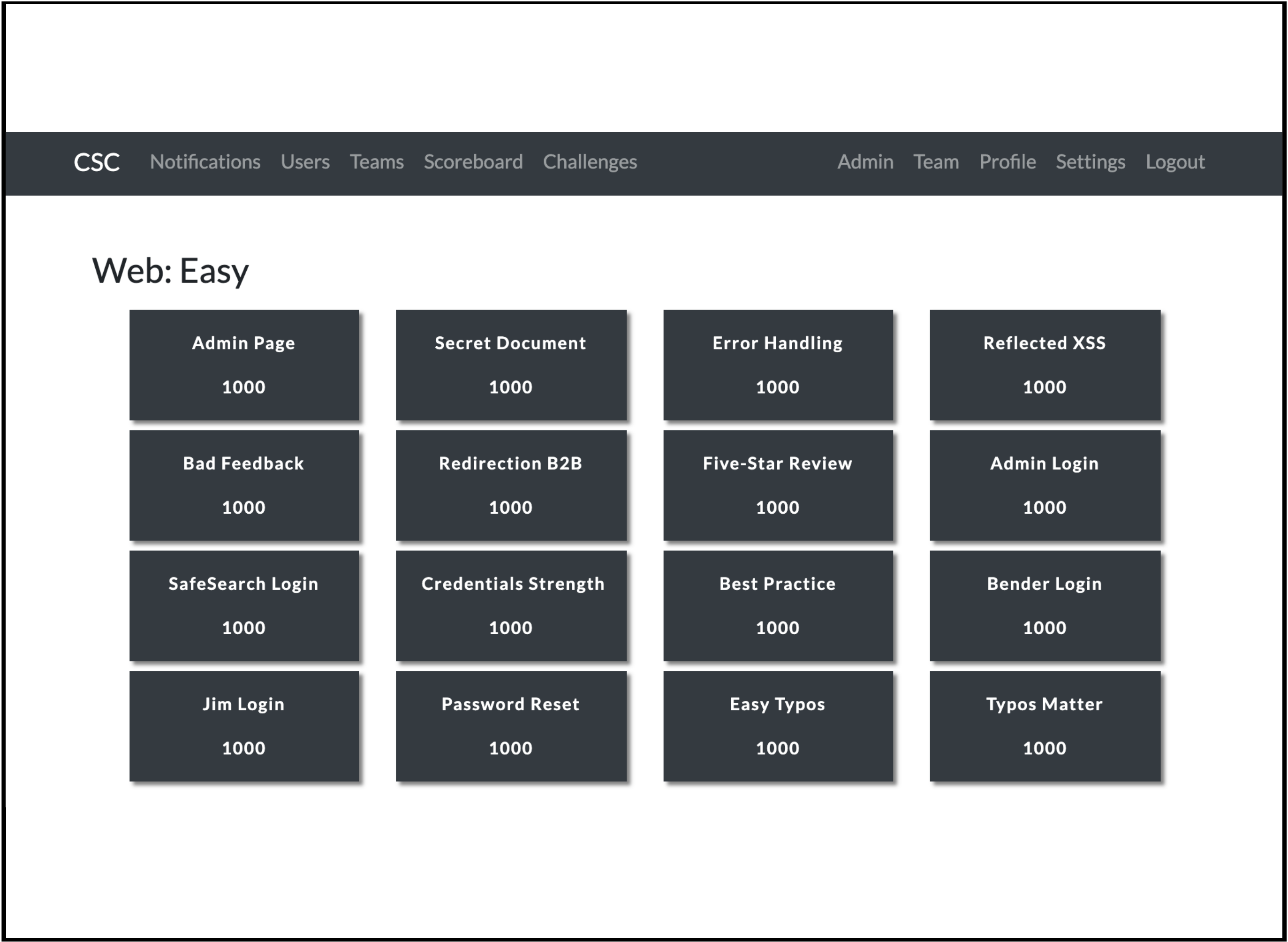}
        \vspace*{-2em}
        \caption{Dashboard}
        \label{fig:dashboard}
        \vspace*{1em}
    \end{minipage}
    \begin{minipage}{.49\textwidth}
        \centering
        \includegraphics[width=.99\linewidth]{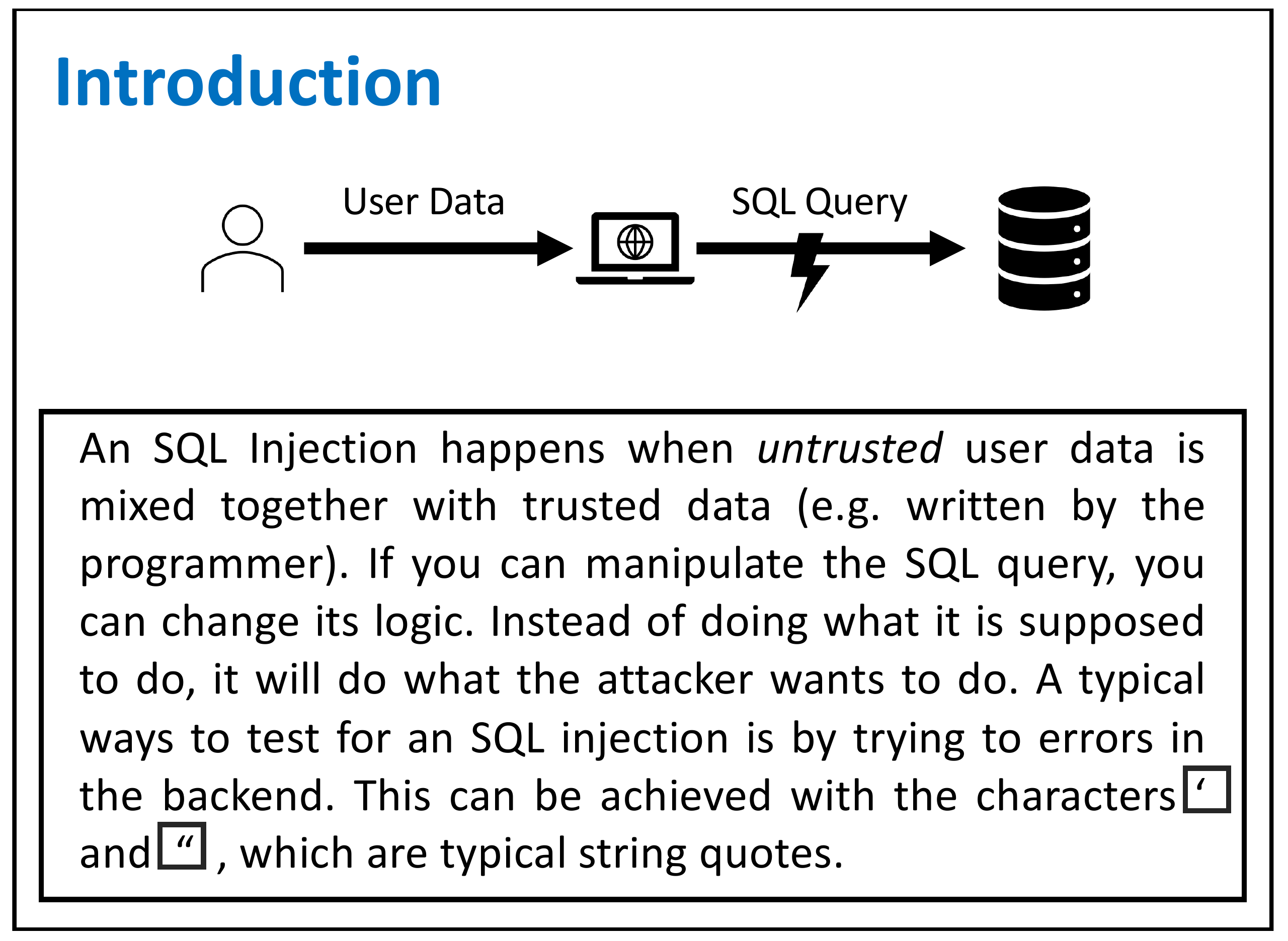}
        \vspace*{-2em}
        \caption{Web Challenge: Phase 1}
        \label{fig:web_challenge_phase_1}
        \vspace*{1em}
    \end{minipage}

    \begin{minipage}{.48\textwidth}
        \centering
        \includegraphics[width=.99\linewidth]{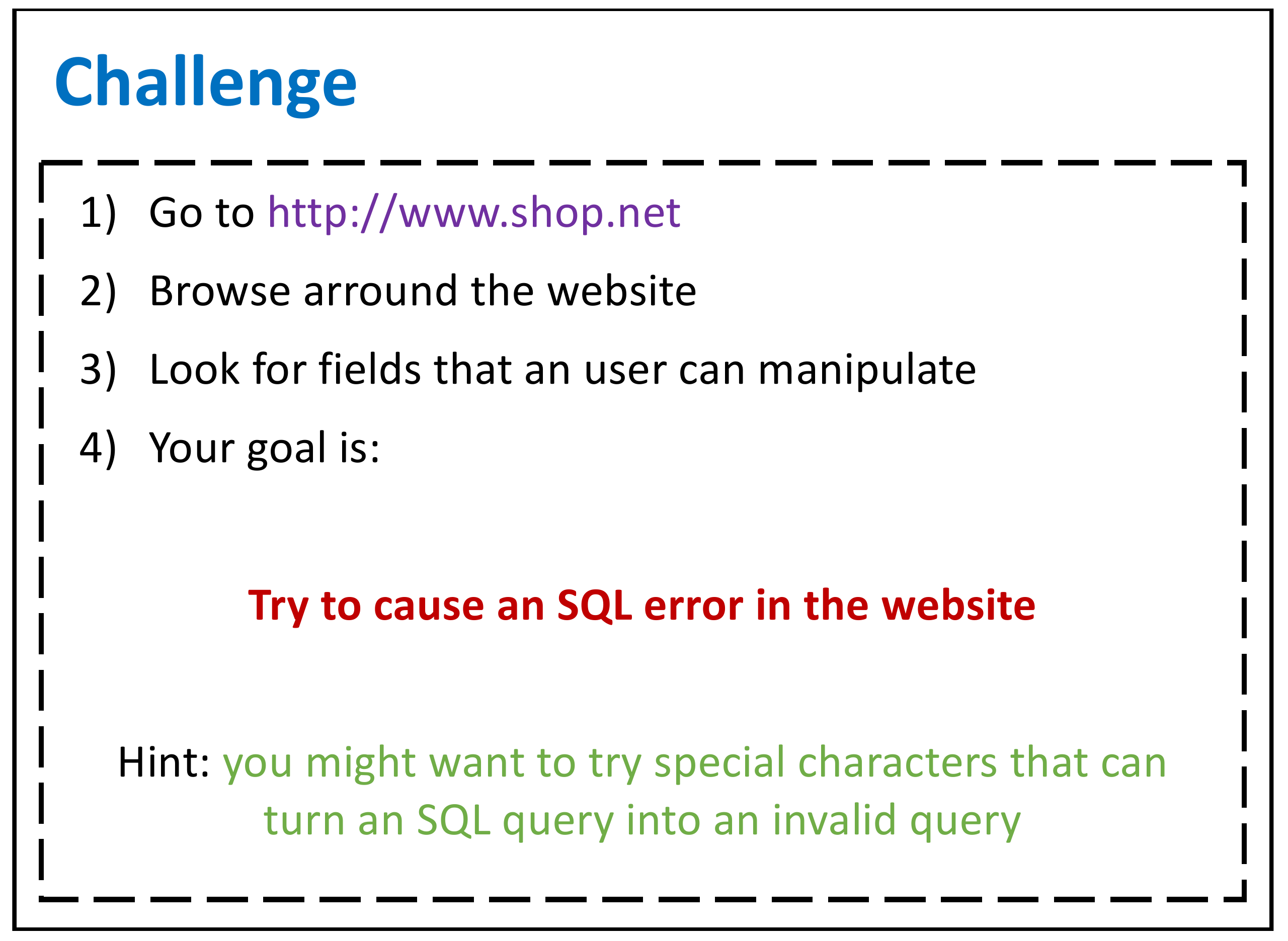}
        \vspace*{-2em}
        \caption{Web Challenge: Phase 2}
        \label{fig:web_challenge_phase_2}
    \end{minipage}
    \begin{minipage}{.48\textwidth}
        \centering
        \includegraphics[width=.99\linewidth]{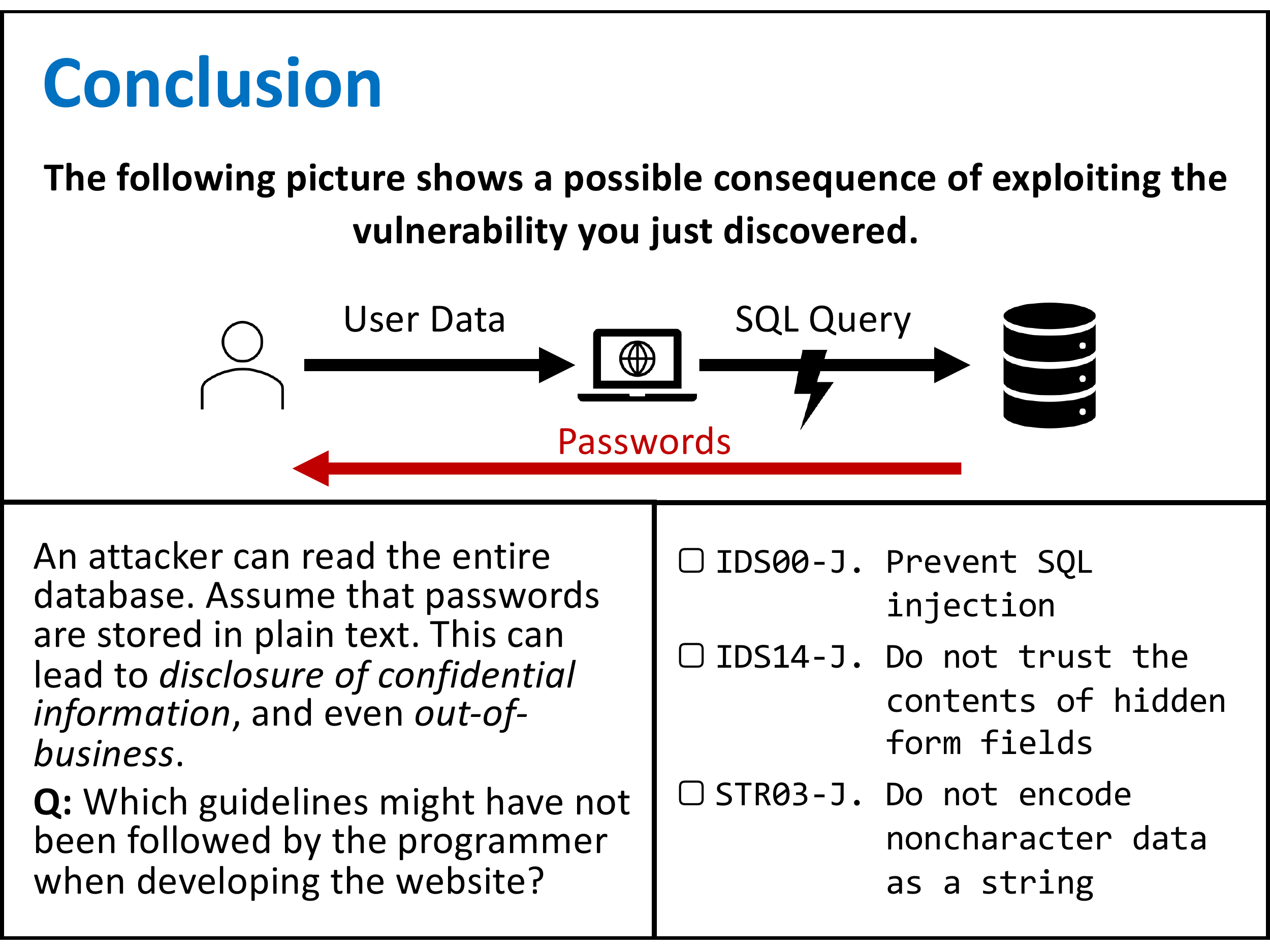}
        \vspace*{-2em}
        \caption{Web Challenge: Phase 3}
        \label{fig:web_challenge_phase_3}
    \end{minipage}
\end{figure}
Challenges on secure coding for software developers can be implemented by using and adapting existing open source components.
Since most of the available projects focus on the offensive perspective, the following adaptations are suggested: 1) include an incomplete description on how to solve the challenge, and 2) provide follow-up questions related to secure coding guidelines.
Fig.~\ref{fig:web_challenge_phase_1}-\ref{fig:web_challenge_phase_3} shows an example of a challenge for Web developers using OWASP JuiceShop.
The challenge's learning goal is to understand what SQL injections are and how to identify an SQL injection quickly.
Phase 1 sets the stage for the challenge (Fig.~\ref{fig:web_challenge_phase_1}).
In Phase 2, the player is assisted with how to find the vulnerability, through the textual description, as in Fig~\ref{fig:web_challenge_phase_2}, or also directed by the game coaches.
The last phase consists of an additional question related to the exercise, as shown in Fig~\ref{fig:web_challenge_phase_3}, which enquires and directs the player to corresponding secure coding guidelines.

Table~\ref{tbl:tools} shows the open-source projects and components which have been used to design CSC challenges for Web and C/C++, along with the expected effort required to modify them.
Note that the design of these challenges is based on open source components that include an offensive perspective. Therefore, after the components' adaptation, it is more natural and accurate to describe these types of challenges as {\it defensive/offensive} (D/O).

\begin{table*}[http]
  \renewcommand{\arraystretch}{0.99}
  \scriptsize
  \centering
  \caption{Open-Source Tools used for Cybersecurity Challenges}
  \label{tbl:tools}
  \vspace{-1em}
  \resizebox{\textwidth}{!}{
  \begin{tabular}{|p{1.4cm}|p{1.8cm}|p{1.1cm}|p{7.5cm}|}
    \hline
    {\bf ~~~Type} & {\bf ~~~~Project} & {\bf ~Effort} & {\bf ~~~~~~~~~~~~~~~~~~~~~~~~~~~Description} \\
    \hline
    \hline
    Web/Java         & Juice Shop               & Minimal & Insecure web application for training purposes from the OWASP project.  \\
    \hline
    Web/Java  & Java & Medium & Secure coding guidelines dedicated to Java from Carnegie Mellon University   \\[-8pt] &SEI-CERT&&\\
    \hline
    Web         & Vulnerable & Medium & REST API containing several vulnerabilities \\[-1pt] & API &&\\
    \hline
    C/C++       & MBE                      & Small  & Vulnerable code from RPISEC course at Rensselaer Polytechnic Institute\\
    \hline
    C/C++       & C/C++           & Medium &Secure coding guidelines dedicated to C/C++ from Carnegie Mellon University    \\[-8pt] &SEI-CERT&&\\
    \hline
    C/C++       & Vulnerable & High & Vulnerable C/C++ code from NIST (Juliet Set) \\[-1pt] & code snippets & & \\
    \hline
  \end{tabular}}
\end{table*}

\subsubsection{Defensive Challenges using Sifu Platform}

\begin{figure}[ht]
    \centering
    \includegraphics[width=\columnwidth]{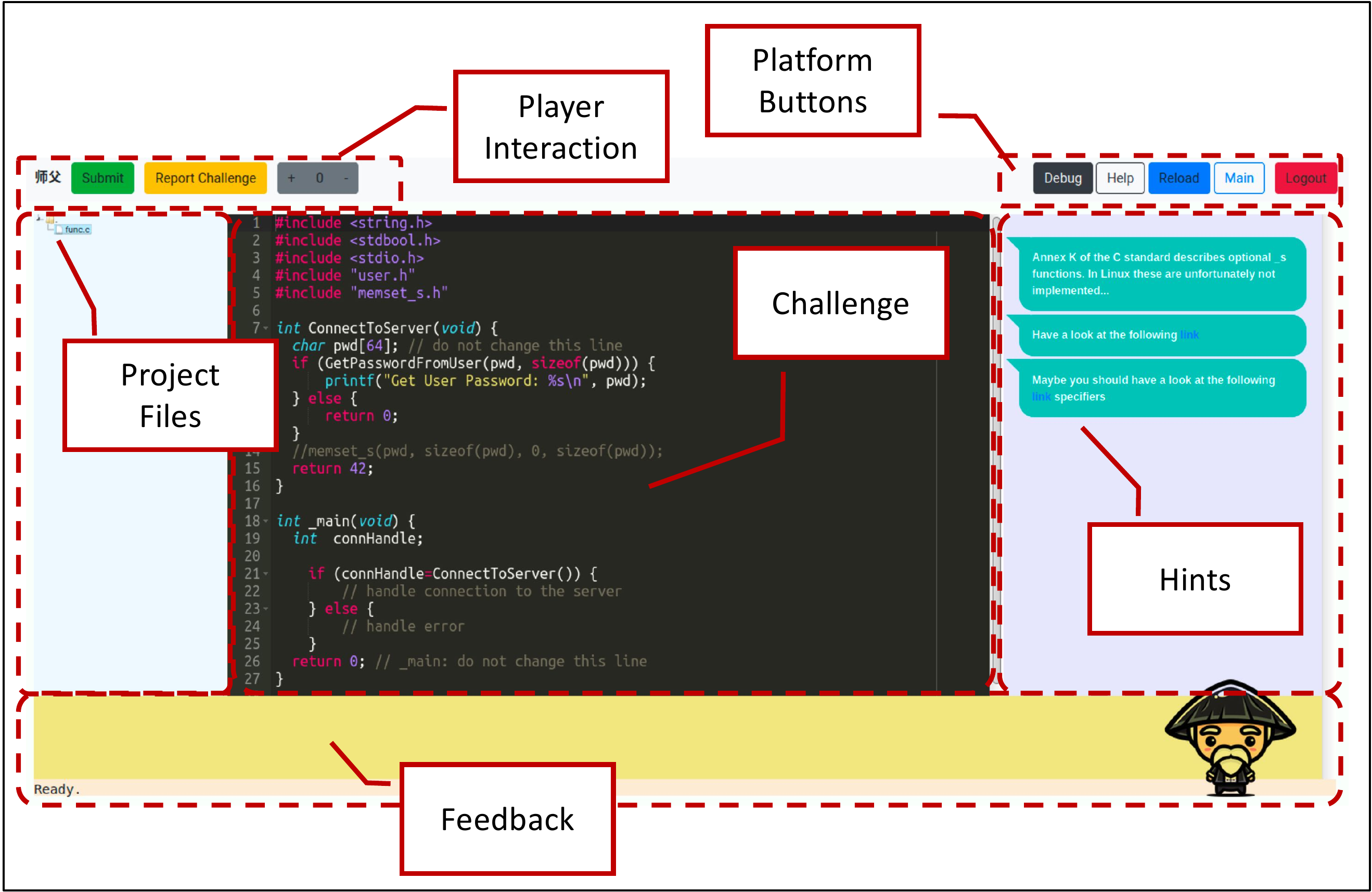}
    \caption{Sifu Platform - User Interface}
    \label{fig:sifu}
\end{figure}

The Sifu platform hosts code projects containing vulnerabilities in a web application.
A web interface is chosen to avoid the players' need to install software on their machines, which might be difficult or impossible in an industrial setting.
The players' task is to fix the project's source code to bring it to an acceptable solution (therefore focusing on the defensive perspective).
An acceptable solution is when the source code is compliant to secure coding guidelines and does not have known vulnerabilities.
The Sifu platform contains two main components: 1) challenge assessment and 2) an automatic coach.
The challenge assessment component analyses the proposed solution submitted by a player and determines if it is acceptable.
Analysis is based on several tools, e.g., compiler output, static code analysis, and dynamic code analysis.
The automatic coach component is implemented through an artificial intelligence technique that provides hints to the participant when the solution is not acceptable, with the intent to guide the participant to an acceptable solution.
Figure~\ref{fig:sifu} shows the web user interface of the Sifu platform.
Note that only phase 2 is shown in the figure.
The player can browse the different files of the project.
All the hints issued by the automatic coach are available on the right-hand side.
If the player experiences errors when using the platform, these can be reported for later analysis and improvement.
Since untrusted and potentially malicious code will be executed in the platform during the analysis stage, several security mechanisms need to be implemented to guarantee that the players cannot hack it.
Further detailed information on the implementation is available in~\cite{Gasiba2020_CyberICPS,Gasiba2020_CyberICPS_Journal}.
The open-source Sifu platform can be downloaded from Github \cite{SifuPlatform}.

\section{Evaluation of CyberSecurity Challenges}

\begin{table}[ht]
\centering
\caption{CyberSecurity Challenge Events}
\label{tbl:events}
\resizebox{\textwidth}{!}{
\begin{tabular}{|c|c|c|c|c|c|c|c|c|c|c|c|c|c|c|c|} 
\hline
 \textbf{No.}   & \textit{\textbf{1}} & \textit{\textbf{2}} & \textit{\textbf{3}} & \textit{\textbf{4}} & \textit{\textbf{5}} & \textit{\textbf{6}} & \textit{\textbf{7}} & \textit{\textbf{8}} & \textit{\textbf{9}} & \textit{\textbf{10}} & \textit{\textbf{11}} & \textit{\textbf{12}} & \textit{\textbf{13}} & \textit{\textbf{14}} & \textit{\textbf{15}}  \\ 
\hline
 \textbf{Type}  & D/O                 & D/O                 & D/O                 & D/O                 & D/O                 & D/O                 & D/O                 & D/O                 & D/O                 & D                    & D                    & D                    & D                    & A                    & A                     \\ 
\hline
 \textbf{Date}  & 11/17               & 5/18                & 7/18                & 7/18                & 9/18                & 7/19                & 7/19                & 9/19                & 10/19               & 6/20                 & 7/20                 & 7/20                 & 7/20                 & 11/20                & 11/20                 \\ 
\hline
 \textbf{NP}    & 11                  & 12                  & 6                   & 30                  & 16                  & 14                  & 15                  & 7                   & 23                  & 15                   & 21                   & 20                   & 15                   & 12                   & 4                     \\ 
\hline
 \textbf{Where} & DE                  & DE                  & DE                  & DE                  & DE                  & CH                  & CH                  & DE                  & TK                  & OL                   & OL                   & OL                   & OL                   & OL                   & OL                    \\
\hline
\end{tabular}
}
 {\\\hspace{\textwidth}
                     {\scriptsize {\bf D/O}: Defensive/Offensive, {\bf D}: Defensive, {\bf NP}: Number of participants, {\bf DE}: Germany, ~~~~~~~~~~~~~~~\\[-4pt] {\bf CH}: China, {\bf TK}: Turkey, {\bf OL}: Online~~~~~~~~~~~~~~~~~~~~~~~~~~~~~~~~~~~~~~~~~~~~~~~~~~~~~~~~~~~~~~~~~~~~~~~~~~~~~} }
\end{table}

The authors have implemented the CSC game and have held a total of thirteen CSC events in the industry: nine onsite events (from November 2017 to October 2019) and four CSC online events (from June 2020 to July 2020). Furthermore, two events in November 2020 were held in the academia. Table \ref{tbl:events} summarizes all the events. 
To evaluate and refine the CSC game, we have performed empirical studies together the CSC events.
The results presented in this  work summarize our empirical studies by focusing on the following six dimensions:
\begin{itemize}
    \item {\bf Know-how} - evaluate if the CSC game contributes to learning new techniques and principles to be used during software development
    \item {\bf Significance} - evaluate if the CSC game contributes to understanding the importance of secure coding guidelines
    \item {\bf Skills} - evaluate if the CSC game contributes to improve the participants' secure coding skills
    \item {\bf Clarity} - evaluate if the challenges in the CSC game are clearly presented
    \item {\bf Coaching} - evaluate if the help provided by coaches is adequate during gameplay
    \item {\bf Behavior} - evaluate if the participants, after playing the CSC game, feel prepared to write secure code
\end{itemize}

The answers to the survey questions were based on a 5-point Likert scale on agreement and are summarized through negative (-) answers (strongly disagree and disagree), neutral (N), and positive (+) answers (agree and strongly agree).
Answering the survey was not mandatory, and the participants that took part in the study have given their consent; additionally, their answers were anonymized.
Although the total number of participants to the CSC events exceeded 200, the total number of participants that answered the survey were: 56 - for defensive/offensive (D/O) events 1-9, 25 - for defensive (D) events 10-13, and 14 - for defensive challenges in the academia (A) in events 14-15.
Additional results were captured through open feedback, questions, and discussions with the participants. The main positive and negative quotes from the participants were also collected.
In the following sub-sections, we present a brief overview and discussion of the survey's main results, participant feedback, and an overview of the lessons learned on the design of CSC games and events.
For a more in-depth overview of the empirical studies, we refer the reader to the work published by the same authors in \cite{gasiba_samra,gasiba_wi,gasiba_re19,Gasiba2019_Raising,Gasiba2020_RankingSCG,Gasiba2020_PlayerProfile,Gasiba2020_CyberICPS,Gasiba2020_CyberICPS_Journal,Gasiba2020_QUATIC,Gasiba2020_TrustCOMM,gasiba_icse}.

\label{sec:results}
\subsection{Results}

\begin{table}
\renewcommand{\arraystretch}{0.9}
\centering
\caption{CyberSecurity Challenge - Empirical Results}
\label{tbl:results}
\resizebox{\textwidth}{!}{
\begin{tabular}{|c|l|c|c|c|l|c|c|c|l|c|c|c|l|l|} 
\cline{1-1}\cline{3-9}\cline{11-13}\cline{15-15}
\multirow{3}{*}{\textbf{Question} } &                      & \multicolumn{7}{c|}{\textbf{Industry} }                                                                                                                        &                      & \multicolumn{3}{c|}{\multirow{2}{*}{\textbf{Academia} }}           &                      & \multicolumn{1}{c|}{\multirow{3}{*}{\textbf{Description } }}                                                           \\ 
\cline{3-9}
                                    &                      & \multicolumn{3}{c|}{\textbf{D/O} }                                 &                      & \multicolumn{3}{c|}{\textbf{D} }                                   &                      & \multicolumn{3}{c|}{}                                              &                      & \multicolumn{1}{c|}{}                                                                                                  \\ 
\cline{3-5}\cline{7-9}\cline{11-13}
                                    &                      & -                    & \textbf{N}           & \textbf{+}           &                      & \textbf{-}           & \textbf{N}           & \textbf{+}           &                      & \textbf{-}           & \textbf{N}           & \textbf{+}           &                      & \multicolumn{1}{c|}{}                                                                                                  \\ 
\cline{1-1}\cline{3-5}\cline{7-9}\cline{11-13}\cline{15-15}
\multicolumn{1}{l}{}                & \multicolumn{1}{l}{} & \multicolumn{1}{l}{} & \multicolumn{1}{l}{} & \multicolumn{1}{l}{} & \multicolumn{1}{l}{} & \multicolumn{1}{l}{} & \multicolumn{1}{l}{} & \multicolumn{1}{l}{} & \multicolumn{1}{l}{} & \multicolumn{1}{l}{} & \multicolumn{1}{l}{} & \multicolumn{1}{l}{} & \multicolumn{1}{l}{} & \multicolumn{1}{l}{}                                                                                                   \\[-8pt]
\cline{1-1}\cline{3-5}\cline{7-9}\cline{11-13}\cline{15-15}
\textit{Q1}                         &                      & 12.5                 & 7.1                  & 80.4                 &                      & 0.0                  & 10.0                 & 90.0                 &                      & 6.2                  & 12.5                 & 81.3                 &                      & \begin{tabular}[c]{@{}l@{}}I learned new techniques and principles\\~ ~ of secure software development\end{tabular}    \\ 
\cline{1-1}\cline{3-5}\cline{7-9}\cline{11-13}\cline{15-15}
\textit{Q2}                         &                      & 0.0                  & 5.3                  & 94.7                 &                      & 0.0                  & 0.0                  & 100.0                &                      & 18.7                 & 12.5                 & 68.8                 &                      & \begin{tabular}[c]{@{}l@{}}I understand the importance of secure\\~ ~ coding guidelines\end{tabular}                   \\ 
\cline{1-1}\cline{3-5}\cline{7-9}\cline{11-13}\cline{15-15}
\textit{Q3}                         &                      & 3.6                  & 14.3                 & 82.1                 &                      & 0.0                  & 0.0                  & 100.0                &                      & 0.0                  & 6.2                  & 93.8                 &                      & \begin{tabular}[c]{@{}l@{}}Focusing on the challenges improves my\\~ ~ practical secure coding skills\end{tabular}     \\ 
\cline{1-1}\cline{3-5}\cline{7-9}\cline{11-13}\cline{15-15}
\textit{Q4}                         &                      & 8.9                  & 8.9                  & 82.2                 &                      & 8.0                  & 8.0                  & 84.0                 &                      & 0.0                  & 12.5                 & 87.5                 &                      & \begin{tabular}[c]{@{}l@{}}The learning goals of the challenges\\~ ~ were clearly explained\end{tabular}               \\ 
\cline{1-1}\cline{3-5}\cline{7-9}\cline{11-13}\cline{15-15}
\textit{Q5}                         &                      & 1.8                  & 12.5                 & 85.7                 &                      & 0.0                  & 0.0                  & 100.0                &                      & 12.5                 & 0.0                  & 87.5                 &                      & \begin{tabular}[c]{@{}l@{}}The help from the coaches was\\~ ~ adequate\end{tabular}                                    \\ 
\hhline{|-~---~---~---~-|}
\textit{Q6}                         &                      & 8.9                  & 26.8                 & 64.3                 &                      & 0.0                  & 20.0                 & 80.0                 &       \multicolumn{1}{c}{}               & \multicolumn{3}{c}{}          &                      & \begin{tabular}[c]{@{}l@{}}I feel that I am prepared to handle\\~ ~ issues related to secure coding at work\end{tabular}  \\
\hhline{|-~---~---~~~~~-|}
\end{tabular}}
{\\\hspace{\textwidth}
                     {\scriptsize {\bf -}: Negative agreement, {\bf N}: Neutral answers, {\bf +}: Positive agreement ~~~~~~~~~~~~~~~~~~~~~~~~~~~~~~~~~~~~~~~~\\[-4pt] {\bf D/O}: Defensive/Offensive, {\bf D}: Defensive ~~~~~~~~~~~~~~~~~~~~~~~~~~~~~~~~~~~~~~~~~~~~~~~~~~~~~~~~~~~~~~~~~~~~~~~~} }
\end{table}

Table \ref{tbl:results} shows a summary of the results for the different six questions, both for the industry (81 participants) and the academia (14 participants).
The two highest-ranked questions are: Defensive/Offensive Challenges - Q2, Q5; Offensive Challenges - Q2+Q3+Q5, Q1; Offensive Challenges - Q3, Q4+Q5.
The results in this table leads to the following conclusions: (1) defensive challenges have a higher level of agreement than defensive/offensive challenges, (2) there is a higher amount of neutral answers in defensive/offensive than in purely defensive challenges, (3) nevertheless both defensive/offensive and defensive challenges show a high level of agreement on the suitability as an method to increase awareness.
These results mean that, while there are good indicators that both challenge types be suitable to raise secure coding awareness on software developers, the indicators for defensive challenges show a higher adequacy.
The presented results also show promising results for the three awareness constructs as introduced by Hänsch et al. \cite{2014_Benenson_Defining_Security_Awareness} - perception (Q2), protection (Q1), and behavior (Q3).
An extended experiment, using the same artifact but in an academic setting, also shows good indicators of its suitability to train future generations of junior industrial software developers.
For a more in-depth discussion on the presented results, we refer the reader to the literature by the same authors \cite{gasiba_samra,gasiba_wi,gasiba_re19,Gasiba2019_Raising,Gasiba2020_RankingSCG,Gasiba2020_PlayerProfile,Gasiba2020_CyberICPS,Gasiba2020_CyberICPS_Journal,Gasiba2020_QUATIC,Gasiba2020_TrustCOMM,gasiba_icse}.

\subsection{Participant Feedback}
Table \ref{tbl:feedback} shows the main positive and negative quotes from participants to the CSC games.
Most of the collected feedback was positive and indicated that the CSC game is suitable for raising secure coding awareness.
The feedback obtained by the authors, during all the events that took place in the industry, has also shown that the software developers highly appreciate playing the CSC game.
For one of the groups that participated in the CSC event, the players have joined forces together after the event and searched the internet for further similar games, thus giving a good indicator of possible long-term effects.
Another success factor was the positive feedback from management, leading to recurring CSC events and establishing good impression managers.
Nevertheless, we collected some negative feedback related to the user interface and the hints' precision.
Additional negative feedback is related to the fact that defensive/offensive challenges still include an offensive part. The offensive part's presence can lead to difficulty in understanding what to do in the challenge due to the participants' background (i.e., software developers).
In a separate discussion, we could conclude that coaches' help can positively improve the game experience.
%Furthermore, we have not obtained this feedback on the defensive challenges, which leads us to conclude that this is not an issue for these types of challenges.

\begin{table}
  \centering
  \renewcommand{\arraystretch}{.9}
  \small
  \caption{Quotes from CSC Participants}
  \label{tbl:feedback}
  \resizebox{\textwidth}{!}{
\begin{tabular}{|c|l|} 
\cline{2-2}
\multicolumn{1}{l|}{}                & \multicolumn{1}{c|}{ \textbf{Quotes from Participants } }                        \\ 
\hhline{-=|}
\multirow{11}{*}{Positive} & I really enjoyed participating in the challenges.                                \\ 
\cline{2-2}
                                     & I am well excited in trying to crack the answers to the challenges               \\ 
\cline{2-2}
                                     & Enjoyed the challenges, different topics and how competitive we became       \\ 
\cline{2-2}
                                     & It was lots of fun. Questions inbetween were nice.                               \\ 
\cline{2-2}
                                     & Enjoyed and lots of fun. I've learned many interesting things                    \\ 
\cline{2-2}
                                     & Quite fun and nice to work, especially work in team                              \\ 
\cline{2-2}
                                     & Enjoyed and learned very much                                                    \\ 
\cline{2-2}
                                     & It was really funny and I leaned a lot                                           \\ 
\cline{2-2}
                                     & Funny and interesting; learned a lot - hope to remember and use in practice  \\ 
\cline{2-2}
                                     & Really liked and enjoyed the exercises                                           \\ 
\cline{2-2}
                                     & Enjoyable to try everything and very fun                                         \\ 
\hline
\hline
\multirow{6}{*}{Negative}  & Hints not always accurate or precisely leading to the problem in the code   \\ 
\cline{2-2}
                                     & We do not perform attacks on systems                                  \\ 
\cline{2-2}
                                     & Could not understand what to do in the challenge                    \\ 
\cline{2-2}
                                     & Some hints are very generic                                                      \\ 
\cline{2-2}
                                     & The user interface is very minimalist                                          \\ 
\cline{2-2}
                                     & User interface could be improved                                                 \\
\hline
\end{tabular}
}
\end{table}

\subsection{Evaluation of the Design}

\begin{figure}[ht]
    \centering
    \includegraphics[width=.99\columnwidth]{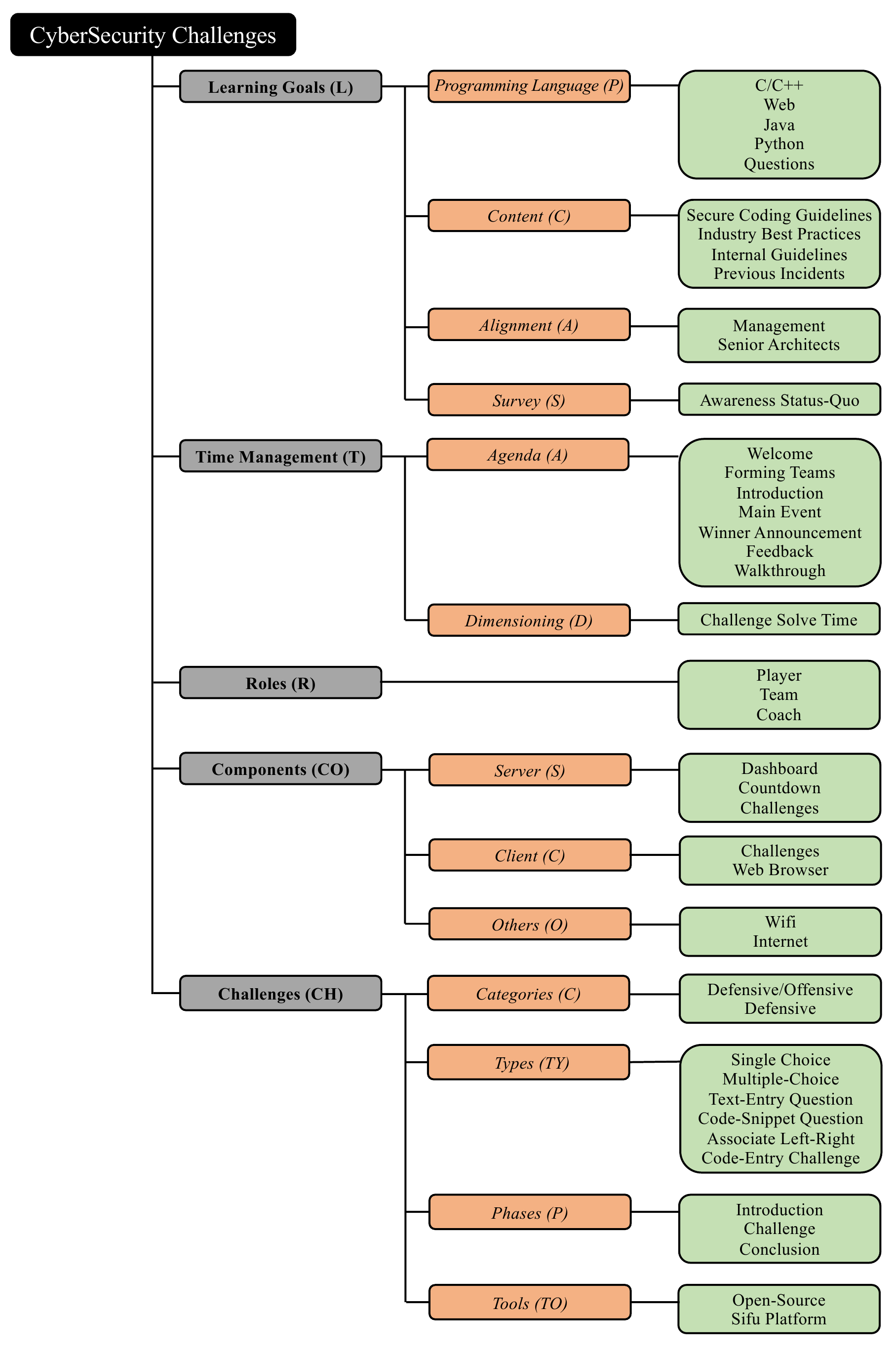}
    \vspace{-1em}
    \caption{CyberSecurity Challenges}
    \label{fig:csc_overview}
\end{figure}
Figure \ref{fig:csc_overview} shows an overview of the lessons learned on the different aspects related with the design, deployment and refinement of CyberSecurity Challenges. These have resulted from all the thirteen deployments that were performed in the industry.
The five top-level design aspects are: 1 - learning goals, 2 - time management, 3 - game roles, 4 - game components, and 5 - challenges.
Learning goals (L) are related to the game's content and adaptation to the target group of software developers and considers programming language, secure coding guidelines, alignment with management, and the current status quo of know-how.
Time management is an essential aspect of deploying and using games in the industry. This aspect includes the agenda of the event and the temporal dimensioning of the challenges.
A clear definition of roles in a serious game is also a critical aspect of such a game's design.
The CyberSecurity Challenges game defines three roles: individual player, team, and coach.
These games are typically deployed in a computer network. Therefore, the different components present in the network and their management are also essential aspects of the game.
Finally, the aspect challenges (CH) looks at the different categories of challenges (as introduced before), challenge types suitable for the industry, the different phases of a challenge, and tools to create the challenges.
Detailed discussions on each of these aspects can be found in \cite{gasiba_samra,gasiba_wi,gasiba_re19,Gasiba2019_Raising,Gasiba2020_RankingSCG,Gasiba2020_PlayerProfile,Gasiba2020_CyberICPS,Gasiba2020_CyberICPS_Journal,Gasiba2020_QUATIC,Gasiba2020_TrustCOMM,gasiba_icse}.

\section{Conclusions}
If not addressed appropriately, software vulnerabilities can result in serious negative consequences. A good time to address these issues is in the early stages of software development by raising the awareness of software developers on secure coding.
This paper presents CyberSecurity Challenges (CSC) as a possible solution.
CyberSecurity Challenges is a genre of serious games developed to raise the awareness of industrial software developers on secure coding and secure coding guidelines.
CSC games have been developed since 2017 in the industry. They were extensively studied as part of the Ph.D. research by the first author, resulting in more than ten publications.
The CSC game can be used both for onsite training and remote training, thus easily adapting to possible travel restrictions imposed by the current COVID-19 situation.

Our results through empirical studies show that this game is adequate to raise secure coding awareness, both when using defensive/offensive challenges and purely defensive challenges.
Furthermore, preliminary results indicate that the same artifact could be used in academia to prepare the future industry workforce.
Feedback obtained from software developers in the industry also indicates this community's acceptance and welcoming of the game.
During gameplay, software developers have fun and practice the usage secure coding guidelines for secure software development.
Furthermore, CSC games found additional success by being well accepted by management.
Therefore, we think that this type of game is a viable approach to tackle possible software vulnerabilities due to bad code quality in terms of security.

\section*{Acknowledgements}
The authors would like to thank the participants of the CyberSecurity Challenges for their time and their valuable answers and comments. The authors would also like to thank Kristian Beckers and Thomas Diefenbach for their helpful, insightful, and constructive comments and discussions.

This resaerch is partly financed by national funds through FCT - Fundação para a Ciência e Tecnologia, I.P., under the projects FCT UIDB/04466/2020 and UIDP/04466/2020. Furthermore, the third author thanks the Instituto Universitário de Lisboa and ISTAR, for their support.

\bibliographystyle{splncs04}
\bibliography{bibliography}

\end{document}